\documentclass[english,twocolumn,journal]{IEEEtran}
\usepackage[T1]{fontenc}
\usepackage{amsmath}
\usepackage{amsthm}
\usepackage{amssymb}
\usepackage{stackrel}
\usepackage{graphicx}
\usepackage{color}

\makeatletter
\theoremstyle{plain}
\newtheorem{thm}{\protect\theoremname}

\IEEEoverridecommandlockouts
\usepackage{ifpdf}
\usepackage{cite}
\ifCLASSOPTIONcompsoc
\usepackage[caption=false,font=normalsize,labelfont=sf,textfont=sf]{subfig}
\else
\usepackage[caption=false,font=footnotesize]{subfig}
\fi
\usepackage{amsmath}

\@ifundefined{showcaptionsetup}{}{
 \PassOptionsToPackage{caption=false}{subfig}}
\usepackage{subfig}
\makeatother

\usepackage{babel}
\providecommand{\theoremname}{Theorem}
\newtheorem{remark}{Remark}

\begin{document}

\title{Millimeter-Wave NOMA Transmission in Cellular M2M Communications for Internet of Things}

\author{Tiejun Lv, \emph{Senior Member, IEEE}, Yuyu Ma, Jie Zeng, \emph{Senior Member, IEEE}, and P. Takis Mathiopoulos, \emph{Senior Member, IEEE}

\thanks{The financial support of the National Natural Science Foundation of China (NSFC) (Grant No. 61671072) is gratefully acknowledged.

T. Lv, Y. Ma and J. Zeng are with the School of Information and Communication Engineering, Beijing University of Posts and Telecommunications (BUPT), Beijing 100876, China (e-mail:
\{lvtiejun, mayy, zengjie\}@bupt.edu.cn).

P. T. Mathiopoulos is with the Department of Informatics and Telecommunications,
National and Kapodistrian University of Athens, Athens 157 84, Greece
(e-mail: mathio@di.uoa.gr).
}}
\maketitle
\begin{abstract}
Massive connectivity and low latency are two important challenges for the Internet of Things (IoT) to achieve the Quality of Service (QoS) provisions required by the numerous devices it is designed to service. Motivated by these challenges, in the paper we introduce a new millimeter-wave non-orthogonal multiple access (mmWave-NOMA) transmission scheme designed for cellular machine-to-machine (M2M) communication systems for IoT applications. It consists of one base station (BS) and numerous multiple machine type communication (MTC) devices operating in a cellular communication environment. We consider its down-link performance and assume that multiple MTC devices share the same communication resources offered by the proposed mmWave-NOMA transmission scheme, which can support massive connectivity. For this system, a novel MTC pairing scheme is introduced the design of which is based upon the distance between the BS and the MTC devices aiming at reducing the system overall overhead for massive connectivity and latency. In particular, we consider three different MTC device pairing schemes, namely i) the random near and the random far MTC devices (RNRF); ii) the nearest near and the nearest far MTC devices (NNNF); and iii) the nearest near and the farthest far MTC device (NNFF). For all three pairing schemes, their performance is analyzed by deriving closed-form expressions of the outage probability and the sum rate. Furthermore, performance comparison studies of the three MTC device pairing schemes have been carried out. The validity of the analytical approach has been verified by means of extensive computer simulations. The obtained performance evaluation results have demonstrated that the proposed cellular M2M communication system employing the mmWave-NOMA transmission scheme improves outage probability as compared to equivalent systems using mmWave with Orthogonal Multiple Access (OMA) schemes.
\end{abstract}


\section{Introduction}
Through the development of numerous applications, the Internet of Things (IoT) \cite{al2015internet,singh2014survey} aims at providing a host of new services to citizens, private and public companies as well as to governmental administrations \cite{zanella2014internet,stankovic2014research,li2018joint}. In general, it is envisioned that the IoT will provide a platform which will connect a huge number of devices in order to gather, share and forward information between devices as well as their users \cite{wunder20145gnow,suomalainen2014smartphone,atzori2010internet}. It is estimated that by the year 2020 almost 50 billion of devices will connected to this platform \cite{liu2016green}. To accommodate the drastically increasing number of these devices, the resulting huge increase in data traffic will have a great impact on the design and implementation of 5th Generation (5G) wireless communication systems. In particular, there will be challenging requirements for their efficient operation, including massive connectivity and low latency \cite{lin2017advanced,liu2017enhancing}. On the one hand, machine-to-machine (M2M) communications have been regarded as one of the promising new technologies to realize IoT employing the 5G network \cite{yang2017energy}. M2M communication systems realize automated data communications among machine type communication (MTC) devices thus constituting the basic communication infrastructure for the emerging IoT \cite{shirvanimoghaddam2016massive,lien2011toward}. In addition, Long Term Evolution (LTE) for MTC (LTE-M) and narrow band IoT (NB-IoT) have been proposed on top of existing cellular standards, which can provide reliable solutions for M2M communications \cite{shirvanimoghaddam2017fundamental}. On the other hand, non-orthogonal multiple access (NOMA), which has been proposed as a multiple access scheme to be employed with 5G wireless communication systems, has the ability to support massive connectivity by means of non-orthogonal resource allocation while simultaneously reducing latencies by its grant-free scheduling. For example, in \cite{higuchi2015non}, an interesting power-domain user multiplexing scheme for future radio access has been proposed. Note that the superposition code (SC) at the transmitter side and successive interference cancellation (SIC) at the receiver side are widely considered in the papers with NOMA transmission \cite{higuchi2015non,higuchi2013non,ding2017noma,liu2017nonorthogonal,kader2017exploiting}.

As opposed to orthogonal multiple access (OMA) schemes, NOMA can support many users via non-orthogonal resource allocation, i.e., multiple users can be served at the simultaneously at the time, frequency and code domains as well as being multiplexed at different power levels \cite{timotheou2015fairness,liu2017fairness,islam2017power}. For example, under poor channel conditions, users are allocated more transmission power as compared to users operating under better channel conditions \cite{ding2016application,liu2017nonOrthogonalMultiple,ding2017asurvey}. Such an approach clearly improves the communication systems' overall fairness. It is noted that since users within one group share the available in the group communication resources, user grouping strategies can significantly influence the overall NOMA system performance \cite{ding2016impact}, and thus it is necessary to carefully study user scheduling schemes \cite{kimy2013non,liu2015user,kim2015uplink,kim2015design}.

Due to the high demand of bandwidth required to support significantly increased data rates, the use of NOMA to millimeter wave (mmWave) becomes a natural choice for 5G systems \cite{zhang2017non,marcano2017performance,ding2017random}. In the past, several studies have been carried out. For example, the authors of \cite{zhang2017non} proposed a cooperative
mmWave-NOMA multicast scheme to improve the mmWave-NOMA multicasting. In \cite{marcano2017performance}, a performance analysis of NOMA in mmWave cells was provided. By considering key features of mmWave systems, such as the high directionality of mmWave transmissions, the performance of the mmWave-NOMA system was analyzed in \cite{ding2017random}. From these and other references it has become clear that mmWave-NOMA transmission has the huge potential for satisfying the specific requirements of cellular M2M communications based IoT.

Device-to-Device (D2D) and M2M communications based on mmWave or NOMA technologies have also attracted considerable attention in both industrial and academic communities \cite{liu2016green,yang2017energy,kong2017millimeter,ding2016mimo,shirvanimoghaddam2016massive}. For example in \cite{liu2016green} the authors have proposed a novel architecture of green relay assisted D2D communications with dual battery for IoT. The capability of mmWave communications for IoT-cloud supported autonomous vehicles was explored in \cite{kong2017millimeter}. A new multiple-input multiple-output (MIMO) NOMA scheme for small packet transmissions in IoT, which was based on some devices that need to be served quickly, was investigated in \cite{ding2016mimo}. The authors in \cite{shirvanimoghaddam2016massive} presented an overview of 3GPP solutions for enabling massive cellular IoT and investigated the random access strategies for M2M communications, which showed that cellular networks should further evolve to support massive connectivity and low latency. In \cite{shirvanimoghaddam2017fundamental}, NOMA was employed to support a large number of devices in cellular systems with limited radio resources. A mmWave-NOMA based relaying scheme was proposed in \cite{sun2017non} aiming at supporting IoT applications. In \cite{yang2017energy,yang2017energyEfficient}, the authors have studied energy-efficient resource allocation for an M2M enabled cellular network. Consequently, the combination of M2M communications and cellular wireless communications is essential for IoT. This combination, in conjunction with the massive connectivity requirements of the IoT, should lead to the use of appropriately modified multiple access techniques. Since in NOMA, a pair of devices share the same communication resource, device pairing can play a key role in improving the performance of NOMA systems. However, to the best of our knowledge, MTC device pairing schemes have not been studied yet in the open technical literature.

Motivated by the above, in this paper we consider a novel mmWave NOMA transmission system for cellular M2M communications tailored for IoT applications. For the efficient operation of the proposed system we effectively pair MTC devices in three schemes according to their distances from the base station (BS), as follows: i) Random near MTC device and random far MTC device (RNRF), i.e. one near MTC device and one far MTC device are randomly selected from two different groups; ii) the nearest near MTC device and the nearest far MTC device (NNNF), in which the nearest nearMTC device and the nearest far MTC device are selected from from two different groups; iii) the nearest near MTC device and the farthest far MTC device (NNFF), in which the nearest nearMTC device and the farthest far MTC device are selected from two different groups.
The main advantages and novelties of the proposed mmWave NOMA scheme can be summarized as follows:
\begin{itemize}
\item Due to the high directionality of mmWave and the excellent collision avoidance of NOMA, the proposed mmWave-NOMA transmission system is capable of achieving massive connectivity in cellular M2M communications. Furthermore, it is shown that by employing random beamforming it is not required from all MTC devices to provide their channel state information (CSI) to the BS, which naturally leads to reduced overhead and latency.
\item Focusing on a single beam, we employ the above mentioned three MTC device pairing schemes which take MTC devices' locations into account in the mmWave-NOMA transmission scheme. These pairing schemes do not require the BS to have knowledge of their CSI, thereby reducing the system overhead. Moreover, transmissions of the MTC devices requiring different channel conditions are easily implemented in NOMA so that quality of service (QoS) requirements of MTC devices can be easily achieved.
\item Closed-form expressions of the outage probability and sum rate at near MTC devices and far MTC devices are derived for the three proposed MTC device pairing schemes in cellular M2M communications employing the mmWave- NOMA transmission scheme. By analyzing the performance of all three MTC device pairing schemes, it is theoretically proven that among the three pairing schemes, NNNF achieves the lowest outage probability both for near MTC and far MTC devices.
\end{itemize}

The rest of this paper is organized as follows: Section II describes the proposed mmWave-NOMA transmission scheme in cellular M2M communications. Section III derives the closed-form expressions of outage probability and sum rate for the proposed MTC device pairing schemes in cellular M2M communications for IoT. Section IV presents various performance evaluation results obtained my means of computer simulations as well as related discussion. Finally, conclusions are provided in Section V.

\section{System Model}
In this section, we first present the channel model used in the considered communication system followed by the detailed description of the proposed transmission scheme. Finally, a detailed derivation of the signal-to-interference-plus-noise ratio (SINR) for the MTC devices will be presented.

\subsection{Channel Model}

Following \cite{lee2016randomly} and \cite{ding2017random}, a typical mmWave channel contains a line-of-sight (LOS) path and several non-line-of-sight (NLOS) paths. Therefore, the mmWave channel vector from the BS to MTC device $k$ can be mathematically modeled as
\begin{equation}
\mathbf{h}_{k}=\sqrt{M}\frac{\alpha_{k,L}\mathbf{a}\left(\theta_{k,L}\right)}{\sqrt{1+d_{k}^{\alpha_{L}}}}+\sqrt{M}\stackrel[l=1]{L}{\sum}\frac{\alpha_{k,NL}\mathbf{a}\left(\theta_{k,NL}^{l}\right)}{\sqrt{1+d_{k}^{\alpha_{NL}}}},
\end{equation}
where $\alpha_{k,L}$ and $\theta_{k,L}$ represent the complex gain and normalized direction of MTC device $k$ for the LOS path, respectively; $\alpha_{k,NL}$ and $\theta_{k,NL}$ represent the complex gain and the normalized direction of MTC device $k$ for the NLOS path, respectively; $L$ is the number of NLOS paths, and $\alpha_{L}$ and $\alpha_{NL}$ are the path loss exponents for the LOS and the NLOS path, respectively; $d_{k}$ denotes the distance from the BS to MTC device $k$. In addition, $\mathbf{a}(\theta)$ is an array steering vector which can be expressed as
\begin{equation}
\mathbf{a}(\theta)=\frac{1}{\sqrt{M}}\left[1,e^{-j\pi\theta},\cdots,e^{-j\pi(M-1)\theta}\right]^{T},\label{eq:2}
\end{equation}
where $\left[\cdot\right]^{T}$ indicates the transpose of matrix.

In mmWave communication systems, the effect of LOS path is dominant because the path loss of NLOS exponents is much larger than that of the LOS exponent, e.g. the power of the signal following the LOS path is $20$ dB higher than the power of the signals following the NLOS paths \cite{lee2016randomly}. Consequently, the dominant path is the LOS path if such path exists, or the dominant path is one of the NLOS paths if a LOS path doesn't exist. Similar to \cite{lee2016randomly} and \cite{ding2017random}, we adopt the single-path (SP) model, so that the mmWave channel simplifies to
\begin{equation}
\mathbf{h}_{k}=\sqrt{M}\frac{\alpha_{k}\mathbf{a}\left(\theta_{k}\right)}{\sqrt{1+d_{k}^{\alpha}}},
\end{equation}
where $\alpha_{k}$ is the complex gain of MTC device $k$ and follows the complex Gaussian distribution with zero mean and variance 1, i.e., $\alpha_{k}\sim\mathcal{CN}\left(0,1\right)$; $\theta_{k}$ is the normalized direction of the dominant path for MTC device $k$, and $\theta_{k}\sim\mathrm{Unif\left[-1,1\right]}$, i.e., $\theta_{k}$ is uniformly distributed between $-1$ and $1$, while $\alpha$ is the path loss exponent.

\subsection{mmWave-NOMA Transmission}

Since conventional beamforming requires that all MTC devices provide their CSI to the BS, system overhead and latency are inevitably increased. In order to reduce them, random beamforming is employed, with each beam servicing two MTC devices. For simplicity, we focus on a single beam, which can be applied to multiple-beam case. The single beam is expressed as
\begin{equation}
\mathbf{p}=\mathbf{a}\left(\nu\right),\label{eq:4}
\end{equation}
which is generated by the BS. In (\ref{eq:4}) and similar to \cite{lee2016randomly} and \cite{ding2017random}, $\nu$ is a random variable with uniformly distributed between $-1$ and $1$, i.e., $\nu\sim\mathrm{Unif\left[-1,1\right]}$. Note that $\mathbf{a}\left(\nu\right)$ is given by (\ref{eq:2}).

According to \cite{lee2016randomly} and \cite{ding2017random}, the effective channel gain of the MTC device $k$, $\left|\mathbf{h}_{k}^{H}\mathbf{p}\right|^{2}$, can be expressed as
\begin{align}
\left|\mathbf{h}_{k}^{H}\mathbf{p}\right|^{2} & =\frac{M\left|\alpha_{k}\right|^{2}\left|\mathbf{a}\left(\theta_{k}\right)^{H}\mathbf{p}\right|^{2}}{1+d_{k}^{\alpha}}=\frac{\left|\alpha_{k}\right|^{2}\left|\stackrel[n=0]{M-1}{\sum}e^{-j\pi n\left(\nu-\theta_{k}\right)}\right|^{2}}{M\left(1+d_{k}^{\alpha}\right)}\nonumber \\
 & =\frac{\left|\alpha_{k}\right|^{2}\sin^{2}\left(\frac{\pi M\left(\nu-\theta_{k}\right)}{2}\right)}{M\left(1+d_{k}^{\alpha}\right)\sin^{2}\left(\frac{\pi\left(\nu-\theta_{k}\right)}{2}\right)}\nonumber \\
 & =\frac{\left|\alpha_{k}\right|^{2}}{\left(1+d_{k}^{\alpha}\right)}F_{M}\left(\nu-\theta_{k}\right),\label{eq:5}
\end{align}
where $F_{M}\left(\cdot\right)$ is the \emph{Fejér kernel}. By increasing $\left(\nu-\theta_{k}\right)$, $F_{M}\left(\nu-\theta_{k}\right)$ goes to zero quickly. If the direction of channel vector of MTC device $k$ aligns to the direction of the beam $\mathbf{p}$, the MTC device will have a large effective channel gain. Furthermore, a large number of MTC devices increase the probability of alignment so that massive connectivity can be more effectively supported by using a mmWave-NOMA transmission scheme.

\begin{figure}[t]
\begin{centering}
\includegraphics[scale=0.5]{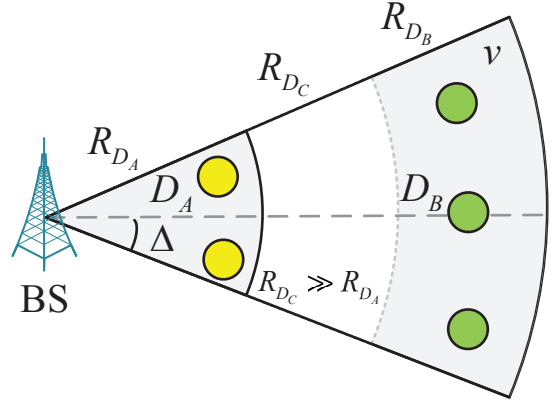}
\par\end{centering}
\caption{The proposed mmWave-NOMA downlink transmission scheme in cellular M2M communications for IoT, which include a BS and two groups of MTC devices, $A=\left\{ A_{i}\right\} $ and $B=\left\{ B_{j}\right\}$ located in the regions $D_{A}$ and $D_{B}$, respectively, which have a central angle of $2\Delta$. Distributions of the near MTC device (yellow circles) and the far MTC device (green circles) follow HPPPs. The MTC devices located in $D_{A}$ and $D_{B}$ will be scheduled. \label{fig:1}}
\end{figure}

In this paper, we introduce a new mmWave-NOMA downlink transmission scheme designed for cellular M2M communications for IoT applications for which one BS serves two groups of MTC devices $A=\left\{ A_{i}\right\} $ and $B=\left\{ B_{j}\right\}$, where $i=1,2,\cdots,N_A$ and $j=1,2,\cdots,N_B$. $N_{k}\left(k\in\left\{ A,B\right\} \right)$ denotes the number of MTC devices in two groups. The BS equipped with $M$ transmit antennas is located at the center of the cell while each MTC device is equipped with a single antenna. As illustrated in Fig. \ref{fig:1}, and according to the operation of the proposed transmission scheme, MTC devices which are located at the wedge-shaped sector $D_{A}$, with an angle of $2\Delta$ and a radius $R_{D_{A}}$, and at the sector ring $D_{B}$ with a maximum radius $R_{D_{B}}$ and a minimum radius $R_{D_{C}}$, are scheduled. It is noted that for the limiting case of $\Delta\rightarrow0$, a large effective channel gain can be achieved.

We consider the scenario in which the MTC devices in group $A$ are deployed within the wedge-shaped sector $D_{A}$, and the devices in group $B$ are deployed within the sector ring $D_{B}$. It is also assumed that $R_{D_{C}}\gg R_{D_{A}}$ so that the channel conditions in these two coverage areas are different for the two groups of MTC devices \cite{liu2016cooperative}. It is further assumed that MTC devices are randomly deployed within the wedge-shaped sector $D_{A}$ and the sector ring $D_{B}$, and that they follow a homogeneous Poisson point process (HPPP) $\Phi_{k}\left(k\in\left\{ A,B\right\} \right)$ with density $\lambda_{k}$. Thus, the probability distribution of $N_{k}\left(k\in\left\{A,B\right\} \right)$ is $P(N_{k}=n)=\mu_{k}^{n}e^{-\mu_{k}}/{n!}$, where $\mu_{A}=\Delta R_{D_{A}}^{2}\lambda_{A}$ and $\mu_{B}=\Delta(R_{D_{B}}^{2}-R_{D_{C}}^{2})\lambda_{B}$.

As previously mentioned, two MTC devices are selected to implement NOMA, with one of them belongs to group $A$ and the other one to group $B$. Furthermore, based on the locations of MTC devices, we consider the following three MTC device pairing schemes to perform NOMA: i) RNRF, in which the near MTC device and the far MTC device are randomly selected from the two groups; ii) NNNF, in which the nearest near MTC device and the nearest far MTC device are selected from the two groups; and iii) NNFF, in which the nearest near MTC device and the farthest far MTC device are selected from the two groups.

\subsection{SINR of MTC Devices}

Let us select one MTC device from each of the two MTC device groups, and the two selected MTC devices are paired to perform NOMA, so that $N_{k}\geq1\left(k\in\left\{ A,B\right\} \right)$. The BS broadcasts the signal $\mathbf{p}\left(\beta_{i1}s_{A_{i}}+\beta_{i2}s_{B_{i}}\right)$ to the near MTC device $A_{i}$ and the far MTC device $B_{i}$, where $s_{A_{i}}$ and $s_{B_{i}}$ are the transmit signals of $A_{i}$ and $B_{i}$,
and $\beta_{i1}$ and $\beta_{i2}$ are their power allocations, respectively, with $\beta_{i1}<\beta_{i2},\ \beta_{i1}^{2}+\beta_{i2}^{2}=1$.

The received signal at the MTC device $A_{i}$ is expressed as
\begin{equation}
y_{A_{i}}=\mathbf{h}_{A_{i}}^{H}\mathbf{p}\left(\beta_{i1}s_{A_{i}}+\beta_{i2}s_{B_{i}}\right)+n_{A_{i}},
\end{equation}
where $n_{A_{i}}$ represents additive white complex Gaussian noise (AWGN).

Considering SIC at the receiver, the MTC device $A_{i}$ first decodes the signal of $B_{i}$, so that the SINR of $B_{i}$ at the receiver of $A_{i}$ can be expressed as
\begin{equation}
\mathrm{SINR}{}_{B_{i}\rightarrow A_{i}}=\frac{\rho\left|\mathbf{h}_{A_{i}}^{H}\mathbf{p}\right|^{2}\beta_{i2}^{2}}{\rho\left|\mathbf{h}_{A_{i}}^{H}\mathbf{p}\right|^{2}\beta_{i1}^{2}+1},\label{eq:7}
\end{equation}
where $\rho$ denotes the transmit signal-to-noise ratio (SNR). Then, $A_{i}$ decodes its own signal, so the SNR of $A_{i}$ is expressed as
\begin{equation}
\mathrm{SINR}{}_{A_{i}}=\rho\left|\mathbf{h}_{A_{i}}^{H}\mathbf{p}\right|^{2}\beta_{i1}^{2}.\label{eq:8}
\end{equation}

Similarly, since the MTC device $B_{i}$ decodes its own signal by treating the signal of MTC device $A_{i}$ as noise, the SINR of $B_{i}$ is expressed as
\begin{equation}
\mathrm{SINR}{}_{B_{i}}=\frac{\rho\left|\mathbf{h}_{B_{i}}^{H}\mathbf{p}\right|^{2}\beta_{i2}^{2}}{\rho\left|\mathbf{h}_{B_{i}}^{H}\mathbf{p}\right|^{2}\beta_{i1}^{2}+1}.
\end{equation}

\section{Performance Analysis of the MTC Device Pairing Schemes}

To guarantee the QoS required by the MTC devices, we define $R_{1}$ and $R_{2}$ as the minimum sum rate of the near MTC device and the far MTC device, respectively, and that $\epsilon_{A_{i}}=2^{R_{1}}-1$ and $\epsilon_{B_{i}}=2^{R_{2}}-1$. When the near MTC device $A_{i}$ cannot decode successfully the signal of the far MTC device $B_{i}$ nor its own signal, outage of the MTC device $A_{i}$ occurs with the following probability
\begin{align}
P_{A_{i}}^{o} & =1-P\left(\mathrm{SINR}{}_{B_{i}\rightarrow A_{i}}>\epsilon_{B_{i}},\mathrm{SINR}{}_{A_{i}}>\epsilon_{A_{i}}\right).\label{eq:10}
\end{align}

Furthermore, the outage probability of MTC device $B_{i}$ is formulated as
\begin{equation}
P_{B_{i}}^{o}=P\left(\mathrm{SINR}{}_{B_{i}}<\epsilon_{B_{i}}\right).\label{eq:11}
\end{equation}

Using (\ref{eq:10}) and (\ref{eq:11}), the outage sum rate of cellular M2M communications with the mmWave-NOMA transmission scheme can be expressed as
\begin{equation}
R_{\mathrm{NOMA}}=\left(1-P_{A_{i}}^{o}\right)R_{A_{i}}+\left(1-P_{B_{i}}^{o}\right)R_{B_{i}},
\end{equation}
while the equivalent outage sum rate of cellular M2M communications with the mmWave-OMA transmission scheme can be expressed as
\begin{equation}
R_{\mathrm{OMA}}=\left(1-P_{A_{i}}\right)R_{A_{i}}^{O}+\left(1-P_{B_{i}}\right)R_{B_{i}}^{O},
\end{equation}
where
\[
P_{n}=P\left(\log\left(1+\rho\left|\mathbf{h}_{n}^{H}\mathbf{p}\right|^{2}\right)<2R_{n}^{O}\right),n\in\left\{ A_{i},B_{i}\right\} ,
\]
and
\begin{equation}\label{eq:osr}
R_{n}^{O}=\frac{1}{2}\log\left(1+\rho\left|\mathbf{h}_{n}^{H}\mathbf{p}\right|^{2}\right),n\in\left\{ A_{i},B_{i}\right\}.
\end{equation}
The reason why the term 1/2 appears in (\ref{eq:osr}) is the fact that the two MTC devices use a resource block, which is shared by two MTC devices in NOMA transmissions \cite{ding2016impact,kimy2013non}.

Next we will analyze the performance of the three MTC device pairing schemes.

\subsection{RNRF Pairing Scheme}

 For this scheme, a near MTC device $A_{i}$ and a far MTC device $B_{i}$ are randomly selected from the two groups with equal probability to be served with the NOMA protocol. It is noted that, since the BS does not require any CSI based on random selection of the MTC devices, the system overhead is significantly reduced.

\subsubsection{Outage Probability of the MTC Near Device of RNRF}

In principle, the outage probability can be obtained by evaluating (\ref{eq:10}) using (\ref{eq:5}), (\ref{eq:7}) and (\ref{eq:8}). However, it is not difficult to realize that this is a very complex task as its solution involves a non-deterministic polynomial-time hard problem. Instead, we will consider the limiting cases for $\Delta\rightarrow0$ and high SNR to obtain the outage probability performance. For this, the following theorem will be used to obtain the outage probability of the near
MTC device of RNRF for arbitrary values of path loss exponent, $\alpha$.

\begin{thm}
For $\Delta\rightarrow0$ and high SNR, the outage probability of the near MTC device $A_{i}$ of RNRF can be approximated as
\begin{equation}
P_{A_{i}}^{o}\approx\frac{\eta_{A_{i}}}{M}\left(2+\frac{\pi^{2}M^{2}\Delta^{2}}{18}\right)\left(\frac{1}{2}+\frac{R_{D_{A}}^{\alpha}}{\alpha+2}\right),\label{eq:15}
\end{equation}
if $\beta_{i2}^{2}-\beta_{i1}^{2}\epsilon_{B_{i}}>0$; otherwise $P_{A_{i}}^{o}=1$.
In the above equation, $\eta_{A_{i}}=\max\left\{ \frac{\epsilon_{B_{i}}}{\rho\left(\beta_{i2}^{2}-\beta_{i1}^{2}\epsilon_{B_{i}}\right)},\frac{\epsilon_{A_{i}}}{\rho\beta_{i1}^{2}}\right\} $.
\end{thm}
\begin{IEEEproof}
$\beta_{i2}^{2}-\beta_{i1}^{2}\epsilon_{B_{i}}\leq0$ indicates the near MTC device cannot decode the signal of the far MTC device successfully, hence $P_{A_{i}}^{o}=1$. When $\beta_{i2}^{2}-\beta_{i1}^{2}\epsilon_{B_{i}}>0$, (\ref{eq:15}) will be derived as follows.

The MTC devices are deployed in $D_{A}$ following HPPPs, so they are independently and identically distributed (i.i.d.) points, denoted by $W_{A_{i}}$, considering the location information $A_{i}$. Therefore, the probability density function (PDF) of $W_{A_{i}}$ can be expressed as
\begin{equation}
f_{W_{A_{i}}}\left(w_{A_{i}}\right)=\frac{\lambda_{A}}{\mu_{A}}=\frac{1}{\Delta R_{D_{A}}^{2}}.\label{eq:16}
\end{equation}

Then, the outage probability of the near MTC device $A_{i}$ is given by
\begin{align}
P_{A_{i}}^{o} & =\int_{D_{A}}\left(1-e^{-\frac{\eta_{A_{i}}\left(1+d_{A_{i}}^{\alpha}\right)}{F_{M}\left(v-\theta_{A_{i}}\right)}}\right)f_{W_{A_{i}}}\left(w_{A_{i}}\right)dw_{A_{i}}\label{eq:17}\\
 & =\frac{1}{\Delta R_{D_{A}}^{2}}\int_{\nu-\Delta}^{\nu+\Delta}\int_{0}^{R_{D_{A}}}(1-e^{-\frac{\eta_{A_{i}}\left(1+r^{\alpha}\right)}{F_{M}\left(v-\theta\right)}})rdrd\theta,\nonumber
\end{align}
where $\eta_{A_{i}}=\max\left\{\frac{\epsilon_{B_{i}}}{\rho\left(\beta_{i2}^{2}-\beta_{i1}^{2}\epsilon_{B_{i}}\right)},\frac{\epsilon_{A_{i}}}{\rho\beta_{i1}^{2}}\right\} $. According to (\ref{eq:5}), the \emph{Fejér kernel} can be written as
\begin{equation}
F_{M}\left(\nu-\theta\right)=\frac{\sin^{2}\left(\frac{\pi M\left(\nu-\theta\right)}{2}\right)}{M\sin^{2}\left(\frac{\pi\left(\nu-\theta\right)}{2}\right)}.
\end{equation}
Noting that $\left|\nu-\theta\right|\leq\Delta$, and following \cite{ding2017random}, for $\Delta\rightarrow0$, the \emph{Fejér kernel} can be approximated as
\begin{align}
F_{M}\left(\nu-\theta\right) & \thickapprox M\mathrm{sinc}^{2}\left(\frac{\pi M\left(\nu-\theta\right)}{2}\right)\nonumber \\
 & \thickapprox M\left(1-\frac{\pi^{2}M^{2}\left(\nu-\theta\right)^{2}}{12}\right).\label{eq:19}
\end{align}
In deriving (\ref{eq:19}) the following approximations have been used: $\sin\left(x\right)\approx x$ for $x\rightarrow0$, $sinc\left(x\right)\approx\left(1-\frac{x^{2}}{6}\right)$ and $\left(1-x\right)^{2}\approx2x$ for $x\rightarrow0$.

Therefore, (\ref{eq:17}) can be approximated as
\begin{align}
P_{A_{i}}^{o} & \approx\int_{\nu-\Delta}^{\nu+\Delta}\int_{0}^{R_{D_{A}}}\frac{1}{\Delta R_{D_{A}}^{2}}\nonumber \\
 & \times\left(1-e^{-\frac{\eta_{A_{i}}\left(1+r^{\alpha}\right)}{M\left(1-\frac{\pi^{2}M^{2}\left(\nu-\theta\right)^{2}}{12}\right)}}\right)rdrd\theta\nonumber \\
 & \approx\int_{\nu-\Delta}^{\nu+\Delta}\int_{0}^{R_{D_{A}}}\frac{1}{\Delta R_{D_{A}}^{2}}\nonumber \\
 & \times\left(1-e^{-\frac{\eta_{A_{i}}\left(1+r^{\alpha}\right)}{M}\left(1+\frac{\pi^{2}M^{2}\left(\nu-\theta\right)^{2}}{12}\right)}\right)rdrd\theta,\label{eq:20}
\end{align}
where the second approximation holds because of $\left(1-x\right)^{-1}\approx\left(1+x\right)$ for $x\rightarrow0$.

Additionally, since $\eta_{A_{i}}$ goes to zero at high SNR, $\left(1-e^{-x}\right)\approx x$ for $x\rightarrow0$ can be used to approximate (\ref{eq:20}) as
\begin{align}
P_{A_{i}}^{o} & \approx\frac{1}{\Delta R_{D_{A}}^{2}}\int_{\nu-\Delta}^{\nu+\Delta}\int_{0}^{R_{D_{A}}}\frac{\eta_{A_{i}}\left(1+r^{\alpha}\right)}{M}\nonumber \\
 & \times\left(1+\frac{\pi^{2}M^{2}\left(\nu-\theta\right)^{2}}{12}\right)rdrd\theta.\label{eq:21}
\end{align}

From (\ref{eq:21}) and after some straightforward mathematical manipulations, (\ref{eq:15}) can be easily derived.
\end{IEEEproof}

\subsubsection{Outage Probability of the Far MTC Device of RNRF}

According to the NOMA principle, the outage of the far MTC device $B_{i}$ appears when it cannot decode its own signal successfully. Again considering the limiting cases for $\Delta\rightarrow0$ and high SNR, the following theorem gives the outage probability of the far MTC device of RNRF for arbitrary values of path loss exponent, $\alpha$.
\begin{thm}
For $\Delta\rightarrow0$ and high SNR, the outage probability of the far MTC device $B_{i}$ of RNRF can be approximated as
\begin{align}
P_{B_{i}}^{o} & \approx\frac{\eta_{B_{i}}}{M\left(R_{D_{B}}^{2}-R_{D_{C}}^{2}\right)}\left(2+\frac{\pi^{2}M^{2}\Delta^{2}}{18}\right)\nonumber \\
 & \times\left(\frac{R_{D_{B}}^{2}-R_{D_{C}}^{2}}{2}+\frac{R_{D_{B}}^{\alpha+2}-R_{D_{C}}^{\alpha+2})}{\alpha+2}\right),\label{eq:22}
\end{align}
if $\beta_{i2}^{2}-\beta_{i1}^{2}\epsilon_{B_{i}}>0$; otherwise $P_{B_{i}}^{o}=1$. In (\ref{eq:22}), $\eta_{B_{i}}=\frac{\epsilon_{B_{i}}}{\rho\left(\beta_{i2}^{2}-\beta_{i1}^{2}\epsilon_{B_{i}}\right)}$.
\end{thm}
\begin{IEEEproof}
Similar to the near MTC device case, the far MTC device cannot decode its own signal successfully when $\beta_{i2}^{2}-\beta_{i1}^{2}\epsilon_{B_{i}}\leq0$, i.e., $P_{B_{i}}^{o}=1$. Next, the outage probability of the far MTC device will be derived when $\beta_{i2}^{2}-\beta_{i1}^{2}\epsilon_{B_{i}}>0$.

Similar to the near MTC device $A_{i}$, the PDF of $W_{B_{i}}$ can be expressed as
\begin{equation}
f_{W_{B_{i}}}\left(w_{B_{i}}\right)=\frac{\lambda_{B}}{\mu_{B}}=\frac{1}{\Delta\left(R_{D_{B}}^{2}-R_{D_{C}}^{2}\right)}.
\end{equation}

Therefore, the outage probability of the far MTC device $B_{i}$ is given by
\begin{align}
P_{B_{i}}^{o} & =\int_{D_{B}}(1-e^{-\frac{\eta_{B_{i}}\left(1+d_{B_{i}}^{\alpha}\right)}{F_{M}\left(v-\theta_{B_{i}}\right)}})f_{W_{B_{i}}}\left(w_{B_{i}}\right)dw_{B_{i}}\label{eq:24}\\
 & =\frac{1}{\Delta(R_{D_{B}}^{2}-R_{D_{C}}^{2})}\int_{\nu-\Delta}^{\nu+\Delta}\int_{R_{D_{C}}}^{R_{D_{B}}}(1-e^{-\frac{\eta_{B_{i}}\left(1+r^{\alpha}\right)}{F_{M}\left(v-\theta\right)}})rdrd\theta,\nonumber
\end{align}
where $\eta_{B_{i}}=\frac{\epsilon_{B_{i}}}{\rho\left(\beta_{i2}^{2}-\beta_{i1}^{2}\epsilon_{B_{i}}\right)}$.

Following a similar procedure as for the near MTC device case, the approximation of (\ref{eq:24}) can be obtained as
\begin{align}
P_{B_{i}}^{o} & \approx\frac{1}{\Delta(R_{D_{B}}^{2}-R_{D_{C}}^{2})}\int_{\nu-\Delta}^{\nu+\Delta}\int_{R_{D_{C}}}^{R_{D_{B}}}\frac{\eta_{B_{i}}\left(1+r^{\alpha}\right)}{M}\nonumber \\
 & \times\left(1+\frac{\pi^{2}M^{2}\left(\nu-\theta\right)^{2}}{12}\right)rdrd\theta.\label{eq:25}
\end{align}

From (\ref{eq:25}) and after some straightforward mathematical manipulations, (\ref{eq:22}) can be easily derived.
\end{IEEEproof}

\subsection{NNNF Pairing Scheme}

 For this scheme, we select a MTC device within the wedge-shaped sector $D_{A}$ which has the shortest distance to the BS as the near MTC device $A_{i^{*}}$. Similarly, we select a MTC device within the sector ring $D_{B}$ which has the shortest distance to the BS as the far MTC device $B_{i^{*}}$. Because of these choices, this scheme can achieve the minimum outage probability of both the near and far MTC devices, which can be considered as an upper bound on the performance. In this case, the BS needs to know only the MTC devices' distance information in NNNF, which leads to a lower system overhead as compared to requiring the knowledge of the MTC devices' effective channel gains.

\subsubsection{Outage Probability of the Near MTC device of NNNF}

Similar to the case of RNRF, the outage of the near MTC device $A_{i^{*}}$ can occur for two reasons. The first one is that the near MTC device $A_{i^{*}}$ cannot decode the signal of the far MTC device $B_{i^{*}}$ successfully, while the second one is that the near MTC device $A_{i^{*}}$ cannot decode its own signal successfully. Based on these, we can analytically obtain the outage probability of the near MTC device of NNNF. The following theorem gives the outage probability of the near MTC device of NNNF for an arbitrary choice of path loss exponent, $\alpha$.
\begin{thm}
For $\Delta\rightarrow0$ and high SNR, the outage probability of the near MTC device $A_{i^{*}}$ of NNNF can be approximated as (\ref{eq:26}) (shown at the top of page 7), if $\beta_{i2}^{2}-\beta_{i1}^{2}\epsilon_{B_{i}}>0$; otherwise $P_{A_{i^{*}}}^{o}=1$. In (\ref{eq:26}), $\gamma\left(\cdot\right)$ denotes the incomplete gamma function.

\begin{figure*}[t]
\begin{align}
P_{A_{i^{*}}}^{o} & \approx\frac{\eta_{A_{i}}\lambda_{A}}{M\left(1-e^{-\Delta\lambda_{A}R_{D_{A}}^{2}}\right)}\left(2\Delta+\frac{\pi^{2}M^{2}\Delta^{3}}{18}\right)\left(\frac{1-e^{-\Delta\lambda_{A}R_{D_{A}}^{2}}}{2\Delta\lambda_{A}}+\frac{\left(\Delta\lambda_{A}\right)^{-\frac{\alpha+2}{2}}}{2}\gamma\left(\frac{\alpha}{2}+1,\Delta\lambda_{A}R_{D_{A}}^{2}\right)\right).\label{eq:26}
\end{align}
\end{figure*}
\end{thm}
\begin{IEEEproof}
The near MTC device cannot decode the signal of the far MTC device successfully when $\beta_{i2}^{2}-\beta_{i1}^{2}\epsilon_{B_{i}}\leq0$, i.e., $P_{A_{i^{*}}}^{o}=1$. Next, the outage probability of the near MTC device will be considered when $\beta_{i2}^{2}-\beta_{i1}^{2}\epsilon_{B_{i}}>0$.

The distance between the nearest $A_{i^{*}}$ and the BS is denoted by $d_{A_{i^{*}}}$. The probability $\Pr\left(d_{A_{i^{*}}}>r\mid N_{A}\geq1\right)$ conditioned on $N_{A}\geq1$ implies that there is no device located in the sector with radius $r$, which is expressed as
\begin{align}
 & \Pr\left(d_{A_{i^{*}}}>r\mid N_{A}\geq1\right)\nonumber \\
= & \frac{\Pr\left(d_{A_{i^{*}}}>r\right)-\Pr\left(d_{A_{i^{*}}}>r,N_{A}=0\right)}{\Pr\left(N_{A}\geq1\right)}\nonumber \\
= & \frac{e^{-\Delta\lambda_{A}r^{2}}-e^{-\Delta\lambda_{A}R_{D_{A}}^{2}}}{1-e^{-\Delta\lambda_{A}R_{D_{A}}^{2}}}.
\end{align}

According to the above expression, the location information about $A_{i^{*}}$ can be obtained. Therefore, the PDF of $d_{A_{i^{*}}}$ is given by
\begin{equation}
f_{d_{A_{i^{*}}}\left(r_{A}\right)}=\frac{2\Delta\lambda_{A}r_{A}}{1-e^{-\Delta\lambda_{A}R_{D_{A}}^{2}}}e^{-\Delta\lambda_{A}r_{A}^{2}}.\label{eq:28}
\end{equation}

Next, the outage probability of the nearest near MTC device $A_{i^{*}}$ is given by
\begin{equation}
P_{A_{i^{*}}}^{o}=\int_{\nu-\Delta}^{\nu+\Delta}\int_{0}^{R_{D_{A}}}\left(1-e^{-\frac{\eta_{A_{i}}\left(1+r^{\alpha}\right)}{F_{M}\left(v-\theta\right)}}\right)\frac{f_{d_{A_{i^{*}}}\left(r\right)}}{2\Delta}drd\theta.\label{eq:29}
\end{equation}

Similar to the near MTC device $A_{i}$ of RNRF, (\ref{eq:29}) can be approximated as
\begin{align}
P_{A_{i^{*}}}^{o} & \approx\int_{\nu-\Delta}^{\nu+\Delta}\int_{0}^{R_{D_{A}}}\frac{\eta_{A_{i}}\left(1+r^{\alpha}\right)}{M}\nonumber \\
 & \times\left(1+\frac{\pi^{2}M^{2}\left(v-\theta\right)^{2}}{12}\right)\frac{f_{d_{A_{i^{*}}}\left(r\right)}}{2\Delta}drd\theta.\label{eq:30}
\end{align}

From (\ref{eq:30}) and after some some straightforward mathematical manipulations, (\ref{eq:26}) can be easily derived.
\end{IEEEproof}

\subsubsection{Outage Probability of the Far MTC device of NNNF}

Similar to the far MTC device of RNRF, the outage of the far MTC device $B_{i^{*}}$ occurs for one situation, namely when the far MTC device $B_{i^{*}}$ cannot decode its own signal successfully. This case characterizes the occurrence of the outage probability for the far MTC which can be obtained for an arbitrary choice of path loss exponent, $\alpha$, through the following theorem.
\begin{thm}
For $\Delta\rightarrow0$ and high SNR, the outage probability of the far MTC device $B_{i^{*}}$ of NNNF can be approximated as (\ref{eq:31}) (shown at the top of page 6) if $\beta_{i2}^{2}-\beta_{i1}^{2}\epsilon_{B_{i}}>0$; otherwise $P_{B_{i^{*}}}^{o}=1$.

\begin{figure*}[t]
\begin{align}
P_{B_{i^{*}}}^{o} &\approx\frac{\eta_{B_{i}}\lambda_{B}}{M\left(1-e^{-\Delta\lambda_{B}\left(R_{D_{B}}^{2}-R_{D_{C}}^{2}\right)}\right)}\left(2\Delta+\frac{\pi^{2}M^{2}\Delta^{3}}{18}\right)e^{\Delta\lambda_{B}R_{D_{C}}^{2}}\nonumber\\
&\times\left(\frac{e^{-\Delta\lambda_{B}R_{D_{C}}^{2}}-e^{-\Delta\lambda_{B}R_{D_{B}}^{2}}}{2\Delta\lambda_{B}}+\frac{\left(\Delta\lambda_{B}\right)^{-\frac{\alpha+2}{2}}}{2}\left(\gamma\left(\frac{\alpha}{2}+1,\Delta\lambda_{B}R_{D_{B}}^{2}\right)-\gamma\left(\frac{\alpha}{2}+1,\Delta\lambda_{B}R_{D_{C}}^{2}\right)\right)\right).\label{eq:31}
\end{align}
\end{figure*}
\end{thm}
\begin{IEEEproof}
The far MTC device cannot decode its own signal successfully when $\beta_{i2}^{2}-\beta_{i1}^{2}\epsilon_{B_{i}}\leq0$, i.e., $P_{B_{i^{*}}}^{o}=1$.
When $\beta_{i2}^{2}-\beta_{i1}^{2}\epsilon_{B_{i}}>0$ the outage probability of the far MTC device will be obtained next.

The distance between the nearest $B_{i^{*}}$ and the BS is denoted by $d_{B_{i^{*}}}$. Similar to (\ref{eq:28}), the PDF of $d_{B_{i^{*}}}$ is expressed as
\begin{equation}
f_{d_{B_{i^{*}}}\left(r_{B}\right)}=\frac{2\Delta\lambda_{B}r_{B}}{1-e^{-\Delta\lambda_{B}\left(R_{D_{B}}^{2}-R_{D_{C}}^{2}\right)}}e^{-\Delta\lambda_{B}\left(r_{B}^{2}-R_{D_{C}}^{2}\right)}.
\end{equation}

Then, the outage probability of the nearest far MTC device $B_{i^{*}}$ is given by
\begin{equation}
P_{B_{i^{*}}}^{o}=\int_{\nu-\Delta}^{\nu+\Delta}\int_{R_{D_{C}}}^{R_{D_{B}}}\left(1-e^{-\frac{\eta_{B_{i}}\left(1+r^{\alpha}\right)}{F_{M}\left(v-\theta\right)}}\right)\frac{f_{d_{B_{i^{*}}}\left(r\right)}}{2\Delta}drd\theta.\label{eq:33}
\end{equation}

Similar to (\ref{eq:21}), (\ref{eq:33}) can be approximated as
\begin{align}
P_{B_{i^{*}}}^{o} & \approx\int_{\nu-\Delta}^{\nu+\Delta}\int_{R_{D_{C}}}^{R_{D_{B}}}\frac{\eta_{B_{i}}\left(1+r^{\alpha}\right)}{M}\nonumber \\
 & \times\left(1+\frac{\pi^{2}M^{2}\left(v-\theta\right)^{2}}{12}\right)\frac{f_{d_{B_{i^{*}}}\left(r\right)}}{2\Delta}drd\theta.\label{eq:34}
\end{align}

From (\ref{eq:34}) and after some straightforward mathematical manipulations, (\ref{eq:31}) can be easily derived.
\end{IEEEproof}

\subsection{NNFF Pairing Scheme}

 For this scheme, we select, within the sector $D_{A}$, a MTC device which has the shortest distance to the BS as the near MTC device $A_{i^{'}}$. Similarly, we select a MTC device within the sector ring $D_{B}$ which has the farthest distance to the BS as the far MTC device $B_{i^{'}}$. If MTC device channel conditions are bigger differences, NOMA can achieve a larger performance gain over OMA, which leads to the NNFF MTC device pairing scheme.

\subsubsection{Outage Probability of the Near MTC device of NNFF}

As for the NNNF case, here also the near MTC device is selected in the same way. In addition, their power allocation factors are identical. Therefore, outage probability of the near MTC device $A_{i^{'}}$ is the same as the outage probability of $A_{i^{*}}$ of NNNF. The approximation of its outage probability expression is given by (\ref{eq:26}), and the proof is the same as that of the Theorem 3.

\subsubsection{Outage Probability of the Far MTC device of NNFF}

Similar to the far MTC device of RNRF, the outage of the far MTC device $B_{i^{'}}$ occurs for one situation, that is the far MTC device $B_{i^{'}}$ cannot decode its own signal successfully. Based on the outage of the far MTC device of NNFF, its outage probability can be obtained for arbitrarily values of $\alpha$, through the following theorem.
\begin{thm}
For $\Delta\rightarrow0$ and high SNR, the outage probability of the far MTC device $B_{i^{'}}$ of NNFF can be approximated as
\begin{align}
P_{B_{i^{'}}}^{o} &\approx\frac{\eta_{B_{i}}\lambda_{B}}{M\left(1-e^{-\Delta\lambda_{B}\left(R_{D_{B}}^{2}-R_{D_{C}}^{2}\right)}\right)}\left(2\Delta+\frac{\pi^{2}M^{2}\Delta^{3}}{18}\right)\nonumber\\
& \times e^{-\Delta\lambda_{B}R_{D_{B}}^{2}}\left(\frac{e^{\Delta\lambda_{B}R_{D_{B}}^{2}}-e^{\Delta\lambda_{B}R_{D_{C}}^{2}}}{2\Delta\lambda_{B}}+\Omega\right),\label{eq:35}
\end{align}
if $\beta_{i2}^{2}-\beta_{i1}^{2}\epsilon_{B_{i}}>0$; otherwise $P_{B_{i^{'}}}^{o}=1$. In (\ref{eq:35}), $\Omega=\int_{R_{D_{C}}}^{R_{D_{B}}}r^{\alpha+1}e^{\Delta\lambda_{B}r^{2}}dr$. \end{thm}
\begin{IEEEproof}
The far MTC device cannot decode its own signal successfully when $\beta_{i2}^{2}-\beta_{i1}^{2}\epsilon_{B_{i}}\leq0$, i.e., $P_{B_{i^{'}}}^{o}=1$.
When $\beta_{i2}^{2}-\beta_{i1}^{2}\epsilon_{B_{i}}>0$, the outage probability of the far MTC device can be obtained as follows.

The distance between the farthest $B_{i^{'}}$ and the BS is denoted as $d_{B_{i^{'}}}$, and the number of MTC devices in $D_{B}$ is denoted as $N_{B}$. Similar to (\ref{eq:28}), the PDF of $d_{B_{i^{'}}}$ can be expressed as
\begin{equation}
f_{d_{B_{i^{'}}}\left(r_{B}\right)}=\frac{2\Delta\lambda_{B}r_{B}}{1-e^{-\Delta\lambda_{B}\left(R_{D_{B}}^{2}-R_{D_{C}}^{2}\right)}}e^{-\Delta\lambda_{B}\left(R_{D_{B}}^{2}-r_{B}^{2}\right)}.
\end{equation}

Then, the outage probability of the farthest far MTC device $B_{i^{'}}$ is given by
\begin{equation}
P_{B_{i^{'}}}^{o}=\int_{\nu-\Delta}^{\nu+\Delta}\int_{R_{D_{C}}}^{R_{D_{B}}}\left(1-e^{-\frac{\eta_{B_{i}}\left(1+r^{\alpha}\right)}{F_{M}\left(v-\theta\right)}}\right)\frac{f_{d_{B_{i^{'}}}\left(r\right)}}{2\Delta}drd\theta.\label{eq:38}
\end{equation}

Similar to (\ref{eq:22}), (\ref{eq:38}) can be approximated as
\begin{align}
P_{B_{i^{'}}}^{o} & \approx\int_{\nu-\Delta}^{\nu+\Delta}\int_{R_{D_{C}}}^{R_{D_{B}}}\frac{\eta_{B_{i}}\left(1+r^{\alpha}\right)}{M}\nonumber \\
 & \times\left(1+\frac{\pi^{2}M^{2}\left(v-\theta\right)^{2}}{12}\right)\frac{f_{d_{B_{i^{'}}}\left(r\right)}}{2\Delta}drd\theta.\label{eq:39}
\end{align}

From (\ref{eq:39}) and after some straightforward mathematical manipulations, (\ref{eq:35}) can be easily derived.
\end{IEEEproof}
Note that when $\alpha$ is a certain value, $\Omega$ has a closed-form expression.

\begin{remark}
For the design of practical IoT systems, if each MTC device requires the same opportunity served and the lowest latency transmission, RNRF should be considered first; if each MTC device requires the best possible performance and low-latency transmission, NNNF should be employed. As far as the NNFF  scheme is concerned, large performance gain can be achieved if MTC device channel conditions are greatly different.
\end{remark}

\subsection{Performance Comparison of the Three Pairing Schemes}

\subsubsection{The Near MTC device}

Compared with (\ref{eq:15}), (\ref{eq:26}) can be rewritten as
\begin{align}
P_{A_{i^{*}}}^{o} & \approx\frac{\eta_{A_{i}}}{M}\left(2+\frac{\pi^{2}M^{2}\Delta^{2}}{18}\right)\left(\frac{1}{2}+L_{A^{*}}\right),
\end{align}
where $L_{A^{*}}=\frac{\Upsilon_{A^{*}}}{2\left(\Delta\lambda_{A}\right)^{\frac{\alpha}{2}}\left(1-e^{-\Delta\lambda_{A}R_{D_{A}}^{2}}\right)}$,
and $\Upsilon_{A^{*}}=\gamma\left(\frac{\alpha}{2}+1,\Delta\lambda_{A}R_{D_{A}}^{2}\right)$ is the incomplete gamma function. When $\Delta\rightarrow0$, $\Upsilon_{A^{*}}$ can be approximated as
\begin{equation}
\Upsilon_{A^{*}}\approx\frac{2\left(\Delta\lambda_{A}\right)^{\frac{\alpha+2}{2}}R_{D_{A}}^{\alpha+2}}{\alpha+2}-\frac{2\left(\Delta\lambda_{A}\right)^{\frac{\alpha+4}{2}}R_{D_{A}}^{\alpha+4}}{\alpha+4},
\end{equation}
which comes from $\left(1-e^{-x}\right)\approx x\ (x\rightarrow0)$, and $\left(1-e^{-\Delta\lambda_{A}R_{D_{A}}^{2}}\right)\approx\Delta\lambda_{A}R_{D_{A}}^{2}$.
Thus, $L_{A^{*}}$ can be approximated as
\begin{equation}
L_{A^{*}}\approx\frac{R_{D_{A}}^{\alpha}}{\alpha+2}-\frac{\Delta\lambda_{A}R_{D_{A}}^{\alpha+2}}{\alpha+4}.
\end{equation}
Obviously, we have $\frac{R_{D_{A}}^{\alpha}}{\alpha+2}<L_{A^{*}}$, which indicates the outage probabilities of the near MTC devices in NNNF and NNFF are less than that of the near MTC devices in RNRF, i.e., $P_{A_{i}}^{o}>P_{A_{i^{*}}}^{o}=P_{A_{i^{'}}}^{o}$.

Consequently, it is clear that the performance of the near MTC devices' outage probability in NNNF equals that of NNFF, and the performance of the near MTC devices' outage probability in RNRF is the worst among the three proposed schemes.

\subsubsection{The Far MTC device}

Similar to the near MTC device, (\ref{eq:31}) can be approximated as
\begin{equation}
P_{B^{*}}^{o}\approx\frac{\eta_{B_{i}}}{M\left(R_{D_{B}}^{2}-R_{D_{C}}^{2}\right)}\left(2+\frac{\pi^{2}M^{2}\Delta^{2}}{18}\right)L_{B^{*}},
\end{equation}
where
\begin{align}
L_{B^{*}} &=e^{\Delta\lambda_{B}R_{D_{C}}^{2}}\Bigg(\frac{e^{-\Delta\lambda_{B}R_{D_{C}}^{2}}-e^{-\Delta\lambda_{B}R_{D_{B}}^{2}}}{2\Delta\lambda_{B}}+\frac{\left(\Delta\lambda_{B}\right)^{-\frac{\alpha+2}{2}}}{2}\nonumber\\
&\times\left(\gamma\left(\frac{\alpha}{2}+1,\Delta\lambda_{B}R_{D_{B}}^{2}\right)-\gamma\left(\frac{\alpha}{2}+1,\Delta\lambda_{B}R_{D_{C}}^{2}\right)\right)\Bigg).\label{eq:43}
\end{align}
When $\Delta\rightarrow0$, (\ref{eq:43}) can be approximated as
\begin{equation}
L_{B^{*}}\approx\frac{R_{D_{B}}^{2}-R_{D_{C}}^{2}}{2}+\frac{R_{D_{B}}^{\alpha+2}-R_{D_{C}}^{\alpha+2}}{\alpha+2}-\Delta\lambda_{B}\frac{R_{D_{B}}^{\alpha+4}-R_{D_{C}}^{\alpha+4}}{\alpha+4}.
\end{equation}
Clearly, $\frac{R_{D_{B}}^{2}-R_{D_{C}}^{2}}{2}+\frac{R_{D_{B}}^{\alpha+2}-R_{D_{C}}^{\alpha+2})}{\alpha+2}>L_{B^{*}}$,
which indicates that the outage probability of the far MTC devices in NNNF is less than that of the far MTC devices in RNRF, i.e., $P_{B_{i}}^{o}>P_{B_{i^{*}}}^{o}$.

Similar to the far MTC device in NNNF, (\ref{eq:35}) can be approximated as
\begin{equation}
P_{B^{'}}^{o}\approx\frac{\eta_{B_{i}}}{M\left(R_{D_{B}}^{2}-R_{D_{C}}^{2}\right)}\left(2+\frac{\pi^{2}M^{2}\Delta^{2}}{18}\right)L_{B^{'}},
\end{equation}
where
\begin{align}
L_{B^{'}} & =e^{-\Delta\lambda_{B}R_{D_{B}}^{2}}\left(\frac{e^{\Delta\lambda_{B}R_{D_{B}}^{2}}-e^{\Delta\lambda_{B}R_{D_{C}}^{2}}}{2\Delta\lambda_{B}}+\Omega\right).
\end{align}
When $\Delta\rightarrow0$, $L_{B^{'}}$ can be approximated as
\begin{equation}
L_{B^{'}}\approx\frac{R_{D_{B}}^{2}-R_{D_{C}}^{2}}{2}+\frac{R_{D_{B}}^{\alpha+2}-R_{D_{C}}^{\alpha+2}}{\alpha+2}+\Delta\lambda_{B}\frac{R_{D_{B}}^{\alpha+4}-R_{D_{C}}^{\alpha+4}}{\alpha+4}.
\end{equation}
In this case, $\frac{R_{D_{B}}^{2}-R_{D_{C}}^{2}}{2}+\frac{R_{D_{B}}^{\alpha+2}-R_{D_{C}}^{\alpha+2})}{\alpha+2}<L_{B^{'}}$,
which indicates the outage probability of the far MTC devices in NNFF is worse than that of the far MTC devices in RNRF, i.e., $P_{B_{i}}^{o}<P_{B_{i^{'}}}^{o}$.

In summary, among the three proposed MTC device pairing schemes, the performance of the far MTC devices' outage probability in NNNF is best, and the performance of the far MTC devices' outage probability in NNFF is worst, i.e., $P_{B_{i^{*}}}^{o}<P_{B_{i}}^{o}<P_{B_{i^{'}}}^{o}$.

\section{Performances Evaluation Results and Discussions}

In this section, various performance evaluation results for the operation of the three proposed MTC device pairing schemes obtained by means of computer simulations complementing the previously derived theoretical approach will be presented. The results obtained for the following system parameter values. The radius of the wedge-shaped sector $D_{A}$ is set as $R_{D_{A}}=\mathrm{2.5\ m}$.  $\lambda_{A}=6$, and $\Delta=0.1$. The radius of the sector ring $D_{B}$ is set as $R_{D_{C}}=\mathrm{8\ m}$ and $R_{D_{B}}=\mathrm{10\ m}$. $\lambda_{B}=2$. The number of  transmit antennas of the BS is $M=4$, and the path loss exponent is set as $\alpha=2$ if there is no other special explanation. $\beta_{i1}^{2}=0.25$ and $\beta_{i2}^{2}=0.75$ are power allocations for the near MTC device and the far MTC device, respectively \cite{liu2016cooperative,ding2017random}. The other parameters are set as $R_{1}=4$ bits per channel use (BPCU) and $R_{2}=1.5$ BPCU. In addition, we focus on LOS path in this paper.

\begin{figure}[t]
\begin{centering}
\subfloat[]{\centering{}\includegraphics[scale=0.32]{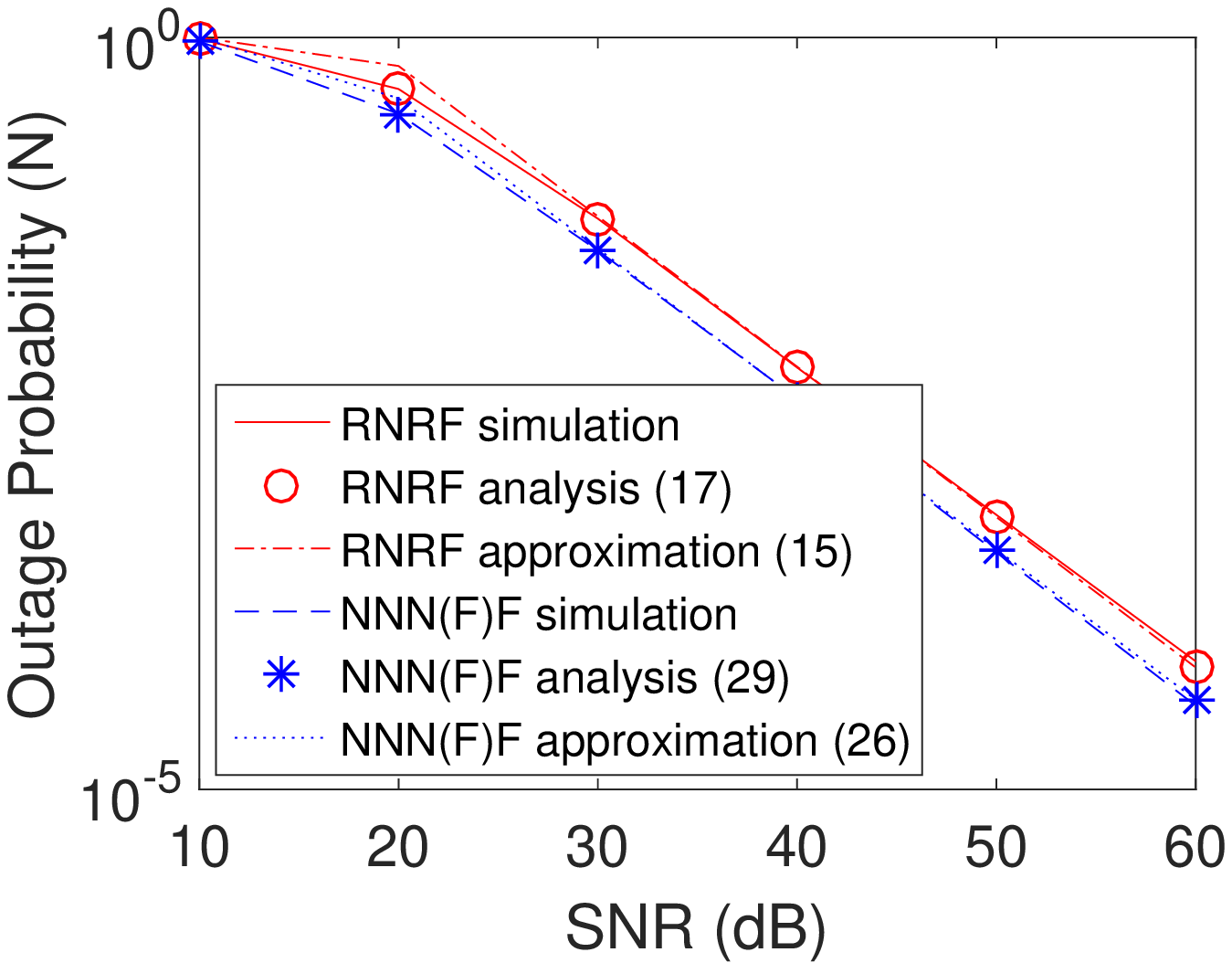}}\subfloat[]{\centering{}\includegraphics[scale=0.32]{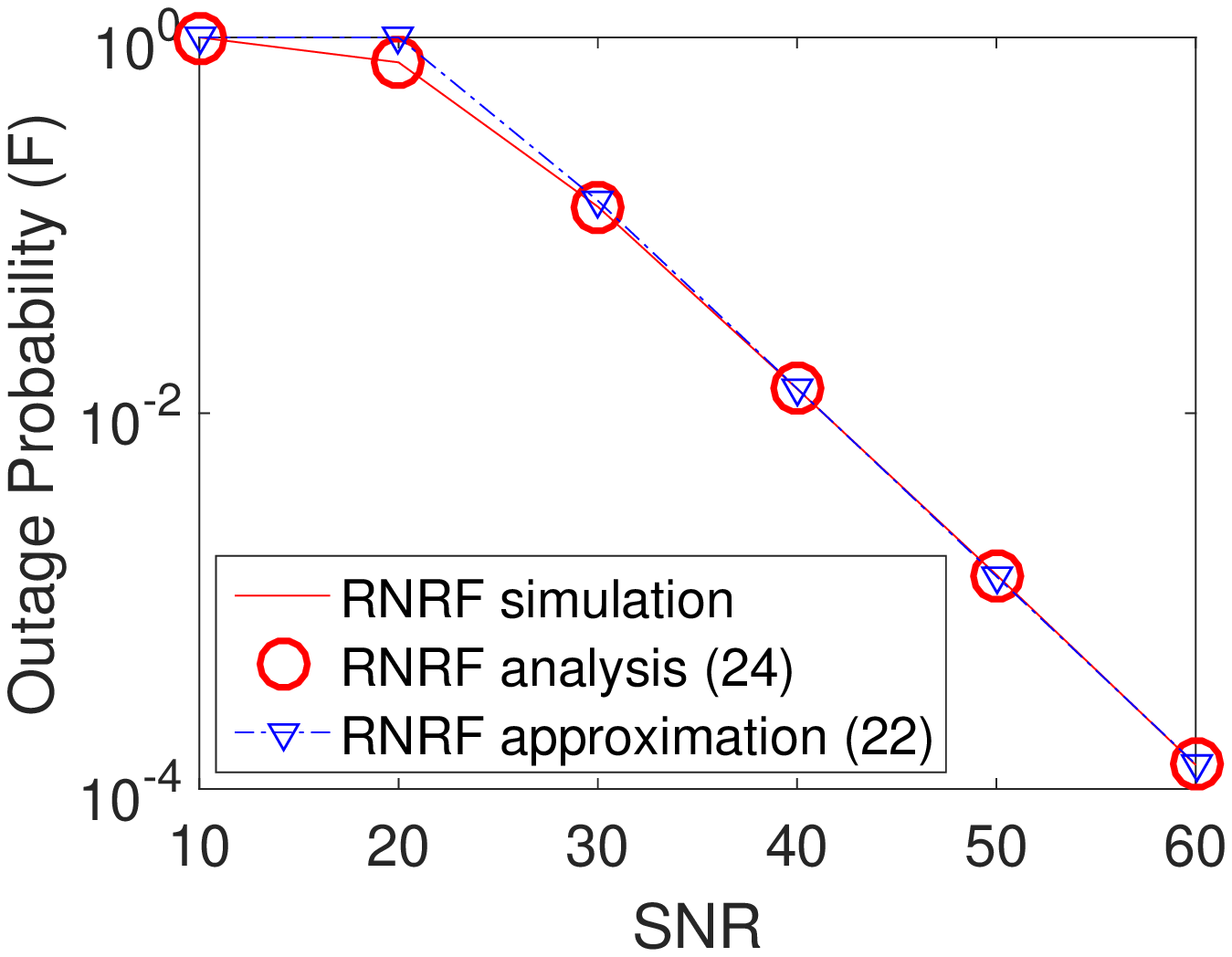}}
\par\end{centering}
\begin{centering}
\subfloat[]{\centering{}\includegraphics[scale=0.32]{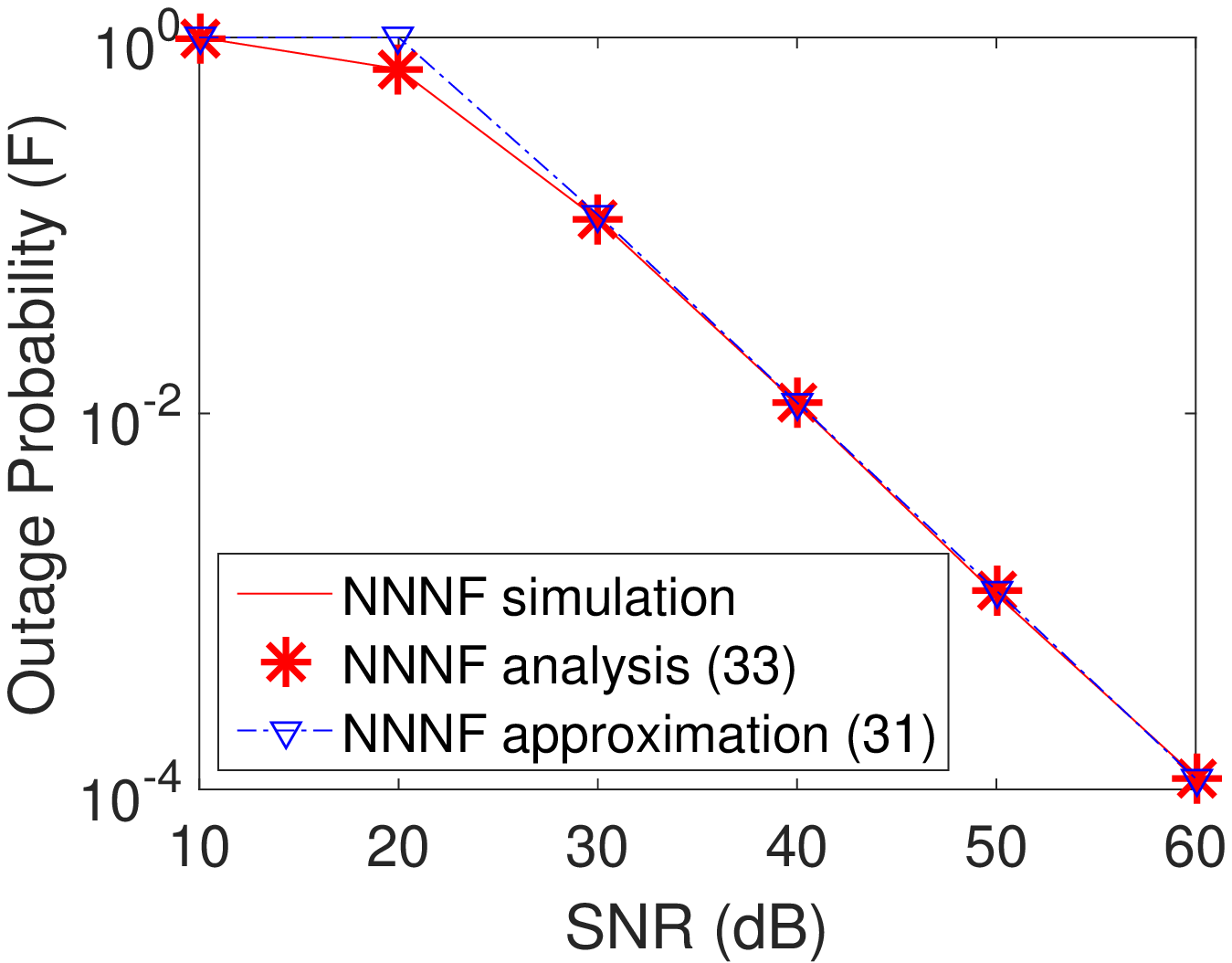}}\subfloat[]{\centering{}\includegraphics[scale=0.32]{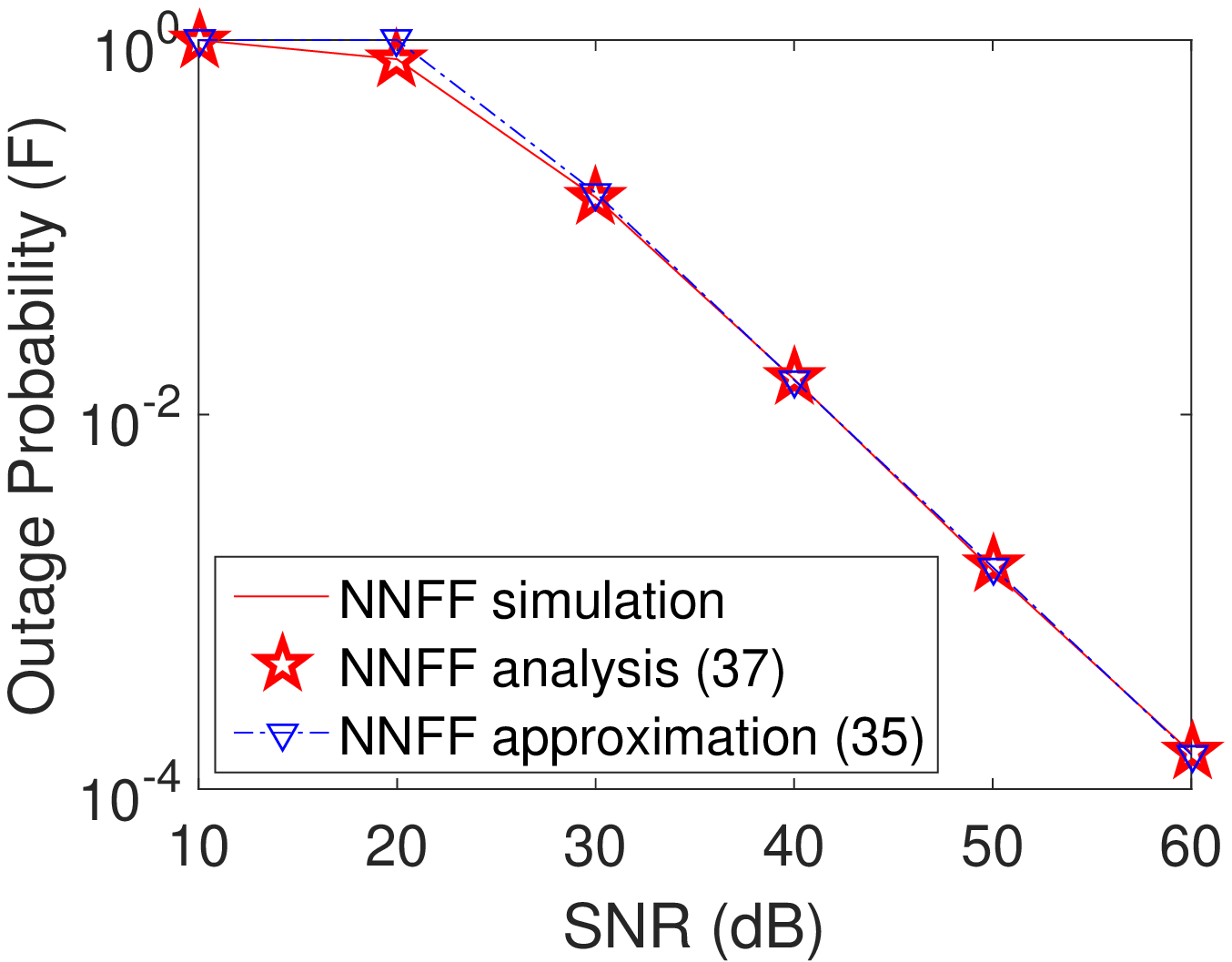}}
\par\end{centering}
\centering{}\caption{Outage probability of MTC devices vs. SNR. (a) the near MTC device in the three MTC device pairing schemes; (b) the far MTC device in RNRF; (c) the far MTC device in NNNF; (d) the far MTC device in NNFF.}
\end{figure}

Fig. 2 plots the outage probability versus SNR. Each subfigure in Fig. 2 includes Monte Carlo simulation results, analytical results and the analytical approximation of outage probability in RNRF, NNNF and NNFF. The outage probability of the near MTC device in NNNF is the same as that of NNFF, which is simplified as NNN(F)F, as shown in Fig. 2 (a). In this figure, the outage probabilities of the near MTC device in the three MTC device pairing schemes are given. Outage probabilities of the far MTC device in RNRF, NNNF and NNFF are presented in Fig. 2 (b), Fig. 2 (c) and Fig. 2 (d), respectively. From these subfigures, the following observations can be made: i) analytical results of RNRF, NNNF and NNFF match the simulation results well; ii) in the high SNR region, the analytical
approximations are very tight; iii) the near MTC device in NNN(F)F achieves a lower outage probability as compared to RNRF.

\begin{figure}[t]
\begin{centering}
\includegraphics[scale=0.6]{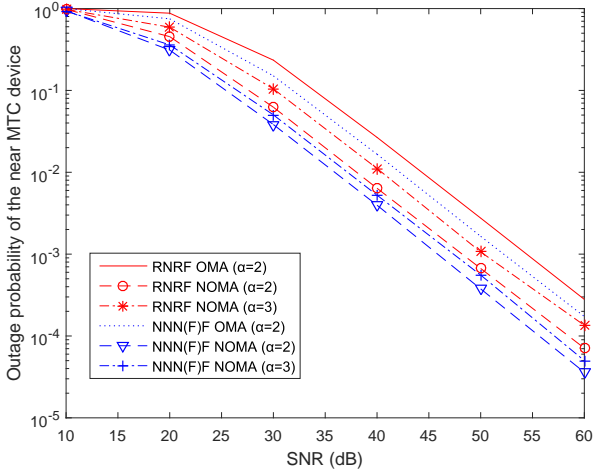}
\par\end{centering}
\caption{Outage probability of the near MTC device vs. SNR for different values of the path loss exponent.}
\end{figure}

Fig. 3 plots the outage probability of the near MTC device versus SNR. The outage probability of the near MTC device versus SNR is given for different values of the path loss exponents of RNRF and NNN(F)F, namely $\alpha=2$ and $\alpha=3$, respectively. From Fig. 3, several observations are obtained as follows: i) the outage probability of the near MTC device in cellular M2M communications with the mmWave-NOMA transmission scheme is better than that with the mmWave-OMA transmission scheme; ii) the outage probability of the near MTC device increases as the path loss exponent increases; iii) among the three schemes, NNN(F)F achieves the lower outage probability; iv) if the outage probability of RNRF is equal to the outage probability of NNN(F)F, the transmit SNR difference between the two schemes is about 3dB.

\begin{figure}[t]
\begin{centering}
\includegraphics[scale=0.6]{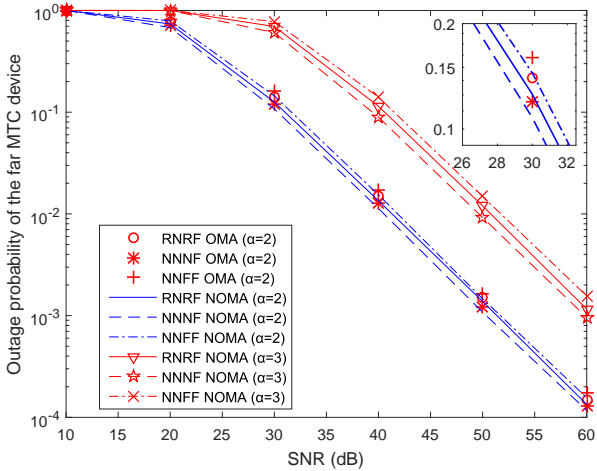}
\par\end{centering}
\caption{Outage probability of the far MTC device vs. SNR for different values of the path loss exponent.}
\end{figure}

Fig. 4 plots the outage probability of the far MTC device versus SNR. The outage probability of the far MTC device versus SNR is given with different path loss exponents of RNRF, NNNF and NNFF. Similar to Fig. 3, the values of the path loss are set as $\alpha=2$ and $\alpha=3$, respectively. From Fig. 4, several observations are obtained as follows: i) the outage probability of the far MTC device in cellular M2M communications with the mmWave-NOMA transmission scheme is better than that with the mmWave-OMA transmission scheme; ii) the outage probability of the far MTC device increases as the path loss exponent increases; iii) among the three schemes, NNNF achieves the lowest outage probability, and NNFF achieves the highest outage probability.

\begin{figure}[t]
\begin{centering}
\subfloat[Monte Carlo simulation results and analytical results of outage sum rates vs. SNR.]{\begin{centering}
\includegraphics[scale=0.6]{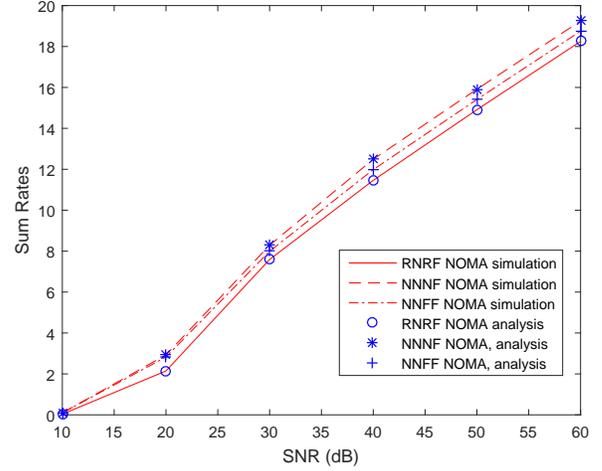}
\par\end{centering}
}
\par\end{centering}
\centering{}\subfloat[Outage sum rates vs. SNR, with different path loss exponents.]{\begin{centering}
\includegraphics[scale=0.6]{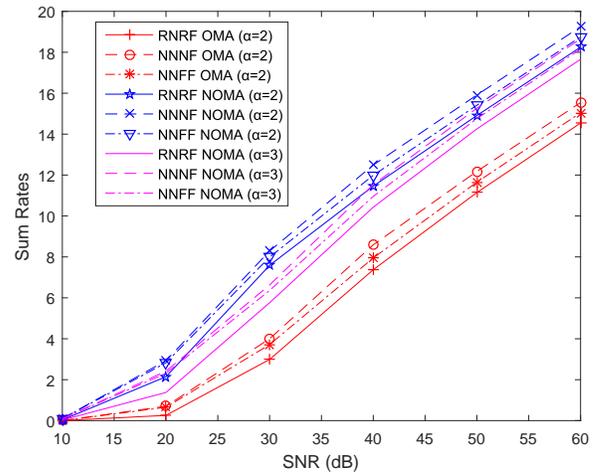}
\par\end{centering}
}\caption{Sum rate of mmWave-NOMA and mmWave-OMA in the proposed MTC device paring schemes vs. SNR.}
\end{figure}

Fig. 5 plots the outage sum rates versus SNR. In Fig. 5 (a), Monte Carlo simulation results and the analytical results of outage sum rates in RNRF, NNNF and NNFF are given. In Fig. 5 (b), outage sum rates under the condition of different SNRs are given with different path loss exponents in the three proposed schemes, and the corresponding OMA simulation results are also given as a benchmark when $\alpha=2$. From Fig. 5, we can observe the following facts: 1) analytical results of RNRF, NNNF and NNFF match the simulation results well; 2) outage sum rates of cellular M2M communications with the mmWave-NOMA transmission scheme are better than that of cellular M2M communications with the mmWave-OMA transmission scheme; 3) outage sum rates of the schemes decrease as path loss exponent increases; 4) among the three proposed schemes, the outage sum rates of the NNNF is best, and the outage sum rates of the RNRF is worst.

\begin{figure}[t]
\centering{}\subfloat[]{\centering{}\includegraphics[scale=0.32]{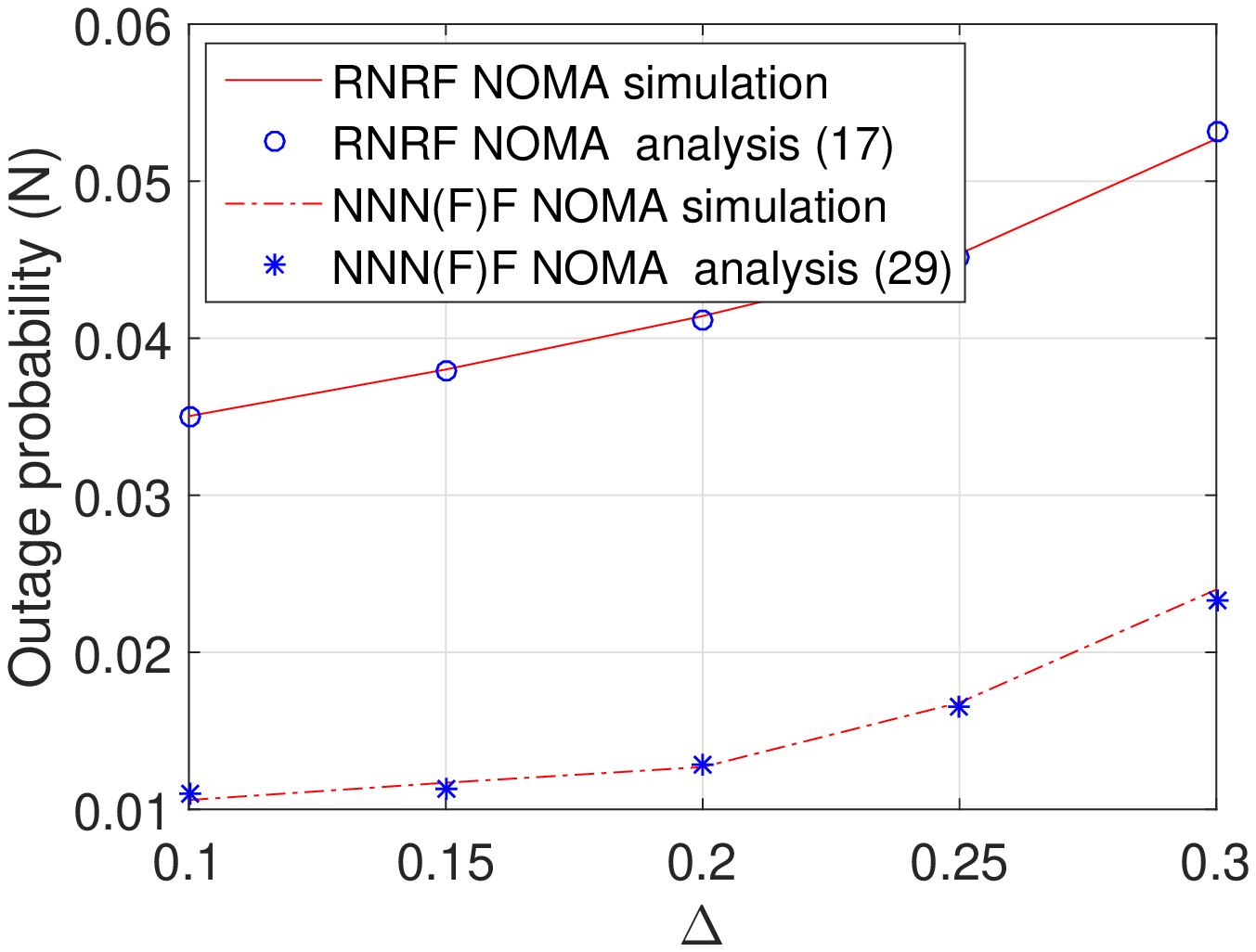}}\subfloat[]{\centering{}\includegraphics[scale=0.32]{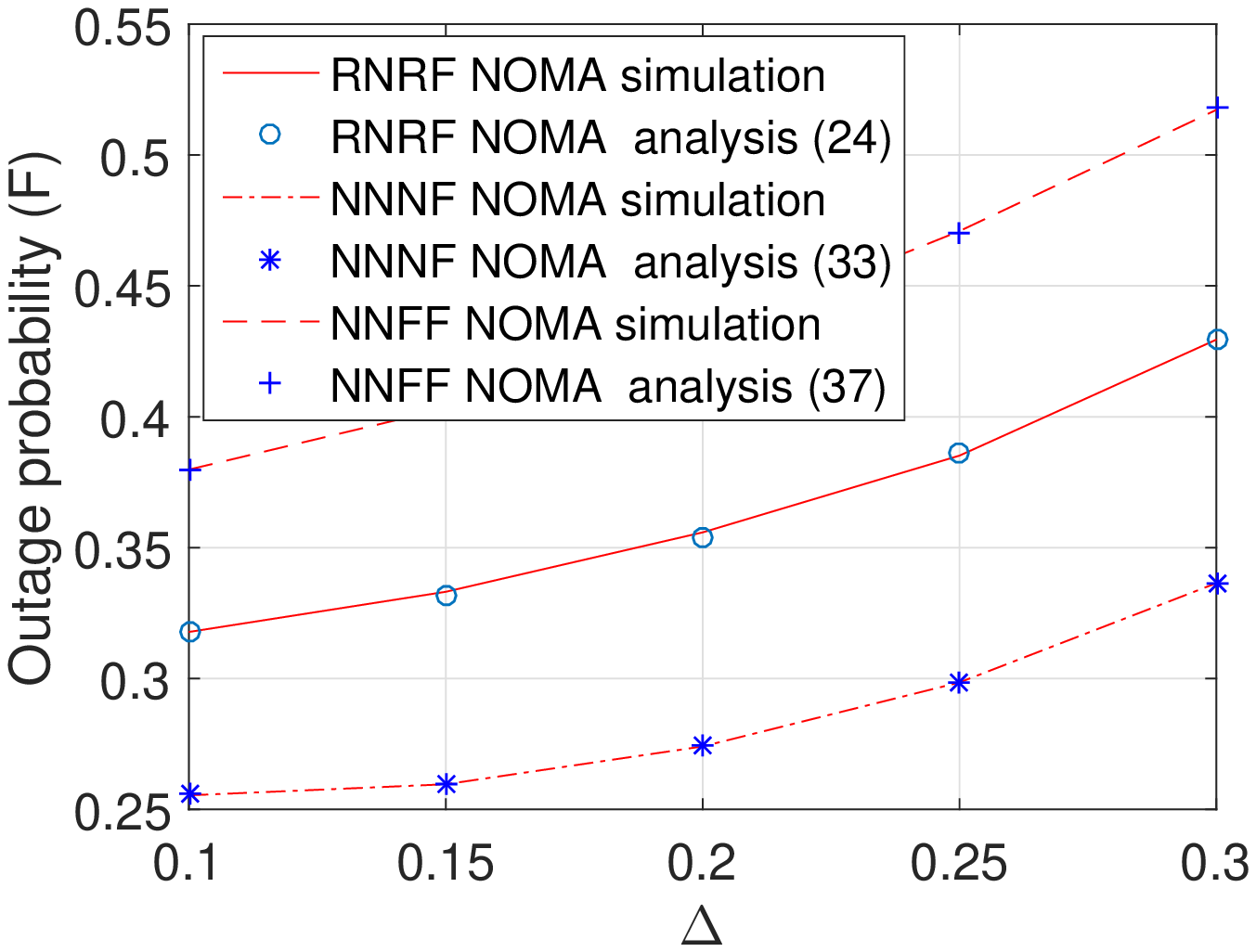}}
\caption{The outage probability vs. $\Delta$. (a) the near MTC device in the three MTC device pairing schemes. (b) the far MTC device in the three MTC device pairing schemes.}
\end{figure}

Fig. 6 plots the outage probability versus $\Delta$. In Fig. 6 (a), the outage probabilities of the near MTC device in the three MTC device pairing schemes are given. In Fig. 6 (b), the outage probabilities of the far MTC device in the three MTC device pairing schemes are shown. From Fig. 6, outage probabilities of the near and far MTC devices increase as $\Delta$ increases, which means that $\Delta\rightarrow0$ can guarantee a large effective channel gain.

\begin{figure}[t]
\begin{centering}
\includegraphics[scale=0.6]{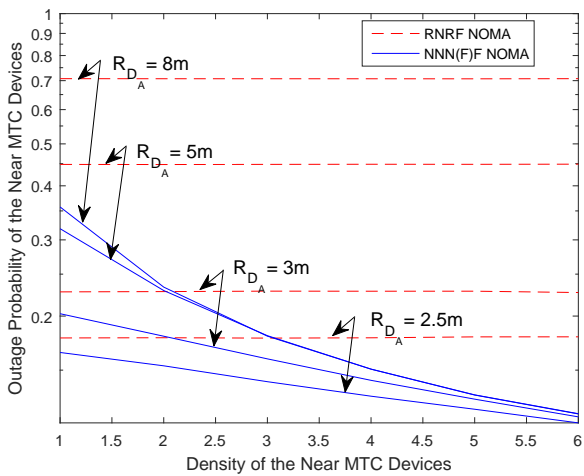}
\par\end{centering}
\caption{The outage probability of the near MTC device vs. density of the near MTC MTC device with different $R_{D_{A}}$, where $R_1=2.5$ BPCU, $R_2=1$ BPCU, $R_{D_{C}}=12$, and $R_{D_{B}}=14$.}
\end{figure}

Fig. 7 plots the outage probability of the near MTC device versus density of the near MTC devices with different $R_{D_{A}}$. The outage probability of the near MTC device in NNN(F)F decrease as the density of the near MTC devices $\lambda_{A}$ increases, because the possibility of scheduling MTC devices with a higher effective channel gain improves. However, outage probability of the near MTC device in RNRF is invariant, this is because that the possibility of scheduling MTC devices with a higher effective channel gain does not change. Furthermore, the outage probability of RNRF and NNN(F)F decreases as $R_{D_{A}}$ decreases, since the path loss of the near MTC devices becomes smaller with the decreasing radius.

\begin{figure}[t]
\begin{centering}
\includegraphics[scale=0.6]{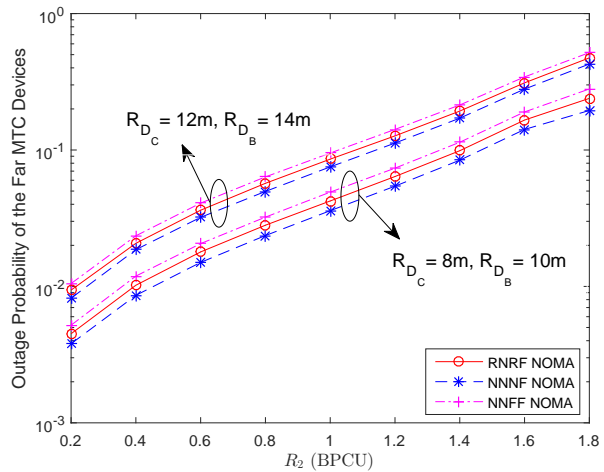}
\par\end{centering}
\caption{The outage probability of the far MTC device vs. $R_2$ with different $R_{D_{C}}$ and $R_{D_{B}}$ in the three pairing schemes.}
\end{figure}

Fig. 8 plots the outage probability of the far MTC device versus $R_2$ with different $R_{D_{C}}$ and $R_{D_{B}}$ in the three proposed pairing schemes. The outage probability of the far MTC device in RNRF, NNNF and NNFF increase as $R_{2}$ increases, this is because QoS of MTC devices becomes higher with the increasing $R_{2}$. Moreover, outage probabilities of RNRF , NNNF and NNFF increase as $R_{D_{C}}$ and $R_{D_{B}}$ increase, since the path loss of the near MTC devices becomes larger with the increasing radius.

\section{Conclusions}

In this paper, a new mmWave-NOMA transmission scheme in cellular M2M communications for IoT which can meet the QoS offered to MTC devices, has been introduced and its performance has been analyzed. Based on the distinct advantages of the proposed mmWave-NOMA transmission scheme, massive connectivity of IoT can be achieved in cellular M2M communications. Using the distance between the MTC device and the BS as a selection criterion, we have proposed three different MTC device pairing schemes which can reduce latency and system overhead, and have focused on a single beam where random beamforming is used. Theoretical studies have shown that among the proposed three schemes, the outage probability of the near MTC device of NNN(F)F is lower than that of the near MTC device of RNRF. Regarding the outage probability of the far MTC device, NNNF and NNFF achieve the best and worst performance respectively. These conclusions have been validated by complementary performance evaluation results obtained by means of Monte Carlo computer simulations.

\bibliographystyle{IEEEtran}
\bibliography{ciations}

\end{document}